\documentclass[]{aastex}
\usepackage[numberedappendix]{emulateapj5,epsfig}

\newlength{\colwidth}

\newcommand{\be}{\begin{equation}}
\newcommand{\ee}{\end{equation}}
\newcommand{\vac}{ \Omega_{\rm v,0} } 
\newcommand{\om}{ \Omega_{\rm m,0} } 
\newcommand{\mob}{ M_{\rm obj} }
\newcommand{\cct}{ \Lambda }
\newcommand{\mpl}{ M_{\rm pl} } 
\newcommand{\bstar}{ \beta^{\star} } 
\newcommand{\rhobar}{\langle \rho \rangle} 
\newcommand{\sig}{ \langle \sigma v \rangle }

\newcommand{\rmax}{ r_{\rm max} } 
\newcommand{\hinv}{\hbox{$h^{-1}$}} 
\newcommand{\msol}{\hbox{${\rm M}_\odot$}} 
\newcommand{\rvir}{\hbox{$r_{\rm 200}$}} 
\newcommand{\mvir}{\hbox{$M_{\rm 200}$}} 
\newcommand{\vvir}{\hbox{$v_{\rm 200}$}} 
\newcommand{\LCDM}{$\Lambda$CDM}
\newcommand{\kmsmpc}{\, \rm{km}\,  \rm{s}^{-1}\, \rm{Mpc}^{-1}}
\newcommand{\lsim}{\lower.5ex\hbox{\ltsima}}
\newcommand{\ltsima}{$\; \buildrel < \over \sim \;$}

\newenvironment{inlinefigure}{\def\@captype{figure}
\noindent\begin{minipage}{0.999\linewidth}\begin{center}}
{\end{center}\end{minipage}\smallskip}

\shortauthors{BUSHA, ADAMS, WECHSLER, \& EVRARD}
\shorttitle{FUTURE EVOLUTION OF STRUCTURE IN AN ACCELERATING UNIVERSE\\} 
\begin{document} 
\submitted{Submitted to the ApJ 2003 April 23}
\journalinfo{Submitted to The Astrophysical Journal}

\title{FUTURE EVOLUTION OF COSMIC STRUCTURE \\ 
IN AN ACCELERATING UNIVERSE} 

\author{Michael T. Busha, Fred C. Adams\altaffilmark{1}, 
Risa H. Wechsler, and August E. Evrard\altaffilmark{1}} 
 
\affil{Michigan Center for Theoretical Physics \\
Physics Department, University of Michigan, Ann Arbor, MI 48109\\
mbusha, fca, wechsler, evrard@umich.edu} 

\altaffiltext{1}{Astronomy Department, University of Michigan, Ann Arbor, MI 48109} 

\begin{abstract} 

Current cosmological data indicate that our universe contains a
substantial component of dark vacuum energy that is driving the cosmos
to accelerate.  We examine the immediate and longer term consequences
of this dark energy (assumed here to have a constant density).  Using
analytic calculations and supporting numerical simulations, we present
criteria for test bodies to remain bound to existing structures.  We
show that collapsed halos become spatially isolated and dynamically
relax to a particular density profile with logarithmic slope steeper
than $-3$ at radii beyond \rvir. The asymptotic form of the space-time
metric is then specified. We develop this scenario further by
determining the effects of the accelerating expansion on the
background radiation fields and individual particles. In an appendix,
we generalize these results to include quintessence.

\end{abstract}

 
\section{INTRODUCTION} 

The past several years have witnessed an impressive solidification of
our estimates for the basic cosmological parameters.  Measurements of
the cosmic microwave background radiation indicate that the cosmos is
spatially flat (e.g., Hanany et al. 2000). Observations of Type Ia
supernovae suggest that the universe is accelerating (Riess et
al. 1998; Perlmutter et al. 1998; Garnavich et al. 1998). These data,
in conjunction with additional observations of the large-scale galaxy
distribution (e.g., Peacock et al. 2001) and the Hubble constant
(Freedman et al. 2001), argue for a specific cosmological model with
matter density $\om \simeq 0.3$, dark vacuum energy density $\vac
\simeq 0.7$, curvature constant $k=0$, and Hubble constant $H_0
\simeq $70 km s$^{-1}$ Mpc$^{-1}$. This cosmological model has
recently received impressive validation by results from the WMAP
satellite (Bennett et al. 2003).  In combination with data on the
Hubble constant, large-scale structure, and they Lyman-$\alpha$
forest, WMAP constrains the values of these parameters to a few
percent (Spergel et al. 2003).

In this newly consolidated cosmological model, the most unexpected
property is the large (apparent) energy density in the form of dark
vacuum energy. The substantial value of $\vac$ = 0.7 produces a
correspondingly large effect on past cosmological evolution. In
particular, the age of the universe is somewhat longer than in a flat
cosmology with no vacuum energy and more consistent with other age
indicators (e.g., Carroll, Press, \& Turner 1992). But, by far, the
most striking consequence of this vacuum energy lies in our
cosmological future.

If the universe is already starting to accelerate, as indicated by the
observationally implied value $\vac$ = 0.7, then structure formation
is virtually finished. In the relatively near future, the universe
will approach a state of exponential expansion and growing
cosmological perturbations will freeze out on all scales. Existing
structures will grow isolated. In the face of such desolation, we
would like to know more quantitatively the conditions required for the
formation (collapse) of future cosmological structures and the
conditions required for small bodies to remain bound to existing
structures. We would also like to know the asymptotic form of the
existing cosmological structures, in particular the dark matter halos.
By answering these questions, we can determine how much structure
formation remains to occur in the future of our universe.

Given the relatively well constrained parameters of our universe, the
future evolution of cosmological structure can now be predicted with
some confidence. Indeed, several recent papers have begun to explore
this issue. Possible future effects of vacuum energy density were
outlined in a recent review of our cosmic future (Adams \& Laughlin
1997, hereafter AL97). As the universe accelerates, currently visible
galaxies are redshifted out of view and will become inaccessible to
future astronomers (Loeb 2002). Simulations of future structure
formation have been done for the case of a cosmological constant with
a focus on our local portion of the universe (Nagamine \& Loeb 2002).
Similar issues have been explored semi-analytically, including basic
effects of quintessence (Chieuh \& He 2002; Gudmundsson \&
Bj{\"o}rnsson 2002). A comprehensive list of papers related to the
future evolution of the universe is compiled in Cirkovic (2003).

In this paper, we consider the future evolution of structure formation
with a constant density of dark vacuum energy.  We extend previous
work by deriving analytic estimates for the conditions required for
the collapse of structures, and by analyzing the results from a suite
of numerical simulations of future structure formation. The future
evolution of cosmic structure can be viewed in two related ways. The
traditional approach considers whether or not a given region of the
universe with overdensity $\delta_0 \equiv (\rho -
\rhobar)/\rhobar$ will collapse. Given the strong suppression
of future structure formation, however, relatively little will happen
in terms of new formation. A related question is to ask whether or not
test bodies will remain bound to existing structures. This issue
operates on a wide range of spatial and mass scales: Will the local
group remain bound to the Virgo cluster? Will a satellite dwarf galaxy
remain bound to the Milky Way?  What bodies would remain bound to an
isolated star in the face of accelerated expansion? This paper
develops both approaches --- analytically in \S 2 and numerically in
\S 3 --- and elucidates the relationship between them.  Our numerical 
simulations also show that the density profiles of dark matter halos
approach a nearly universal form.  Every dark matter halo grows
asymptotically isolated and becomes the center of its own island
universe; each of these regions of space-time then approaches a
universal geometry. In \S 4, we determine the sphere of influence of
existing structures and find the future time at which they grow
cosmologically isolated. Next, we consider the implications of cosmic
acceleration for the background radiation fields, suppression of
particle annihilation, and the long term geometry of space-time.  
Because the equation of state of the dark vacuum energy is not
completely determined, we generalize these results (in an Appendix)
to consider dark energy that varies with time (equivalently, the
scale factor).

\section{ANALYTICAL DESCRIPTIONS OF STRUCTURE FORMATION} 

Before performing detailed numerical simulations of future structure
formation, it is useful to develop simple analytical estimates.
Such results have been developed previously to account for past
structure formation in an accelerating universe (see especially Lokas
\& Hoffman 2001). While most past work has emphasized the evolution of
structure up to the present epoch, we focus here on its evolution into
the future. Throughout this treatment, we assume a spatially flat
universe, which at the present epoch has $\om$ = 0.3 and $\vac$ = 0.7 
($\om + \vac$ = 1).  In the future, the values of $\Omega_{\rm m}$
and $\Omega_{\rm v}$ vary, but their sum continues to equal unity.

\subsection{Collapse of overdense regions} 

The evolution of a given region of the universe is described by an 
energy equation. For a spherical patch of physical size $r$, this energy 
equation can be written in the form 
\be
{1\over 2} {\dot r}^2 - {GM \over r} - {1 \over 2} H^2 r^2 \vac = E \, , 
\ee
where the energy $E$ is given by 
\be
E = {1 \over 2} H_0^2 r_0^2 \bigl[ 1 - \om (1 + \delta_0) - \vac 
\bigr] \, . 
\ee
This set of equations implicitly assumes that ${\dot r} = H_0 r_0$,
i.e., that the particles are traveling along with the unperturbed
Hubble flow at the present epoch. Later in this section, we generalize
this analysis to consider the case in which particles have already
slowed down relative to the Hubble flow due to the past action of
gravity.

If we define the dimensionless variable $\xi \equiv r/r_0$ and the
dimensionless time $\tau \equiv H_0 t$, then the energy equation can
be rewritten in the simpler form 
\be
\bigl( {d\xi \over d\tau} \bigr)^2 = 1 + \om
\bigl( {1\over\xi} - 1 \bigr) (1 + \delta_0) + \vac (\xi^2 - 1) \, . 
\ee
If a cosmological structure is slated to collapse in the future, then
the effective velocity (the time derivative of $\xi$) must change
sign, which requires the right hand side of the above equation to
vanish. This requirement results in a cubic equation of the form 
\be
\xi^3 + \bigl[ {1 \over \vac} - 1 - {\om \over \vac} (1 + \delta_0) 
\bigr]\xi + {\om \over \vac} (1 + \delta_0) = 0 \, . 
\ee 
For a given (flat) cosmology (a given value of $\om$, which in turn 
specifies $\vac = 1 - \om$), and a given overdensity $\delta_0$, 
the cubic has three real roots if the following constraint is 
satisfied: 
\be
\bigl[ {\om \over \vac} (1 + \delta_0) \bigr]^ 2 < {4 \over 27} 
\bigl[ 1 + {\om \over \vac} (1 + \delta_0) - {1 \over \vac} \bigr]^3 \, . 
\ee
The minimum overdensity $\delta_0$ occurs when the above constraint 
is saturated (i.e., at equality). For a given value of $\om$, the 
vacuum energy density $\vac = 1 - \om$ is specified, and the equation 
has a given root. For the currently favored cosmological model with 
$\om$ = 0.3, the root occurs for $\delta_0$ = 17.6. 

\subsection{Criterion for being bound to existing structures}

A related issue is to ask whether a small mass or test particle will
be bound to currently existing cosmological structures. To carry out
this calculation, we consider an existing object of mass $\mob$, which
could be (the dark matter halo encompassing) 
a galaxy or a cluster of galaxies. We then ask whether test particles
--- much smaller structures exterior to the system --- will be bound to
$\mob$ or not. The test particles start with a radial distance $r_0$
at the present epoch. For simplicity, we assume that the distance
$r_0$ is much larger than the size of the collapsed object so that the
potential of a point mass provides a good approximation.

The calculation is analogous to that of the previous section, with 
the mass of the potentially collapsing system replaced by the mass 
$\mob$ of the pre-existing structure. The energy equation can again 
be written in non-dimensional form 
\be
\bigl( {d\xi \over d\tau} \bigr)^2 = 
1 + (\om + \beta) \bigl[ {1 \over \xi} - 1 \bigr] + \vac 
(\xi^2 - 1) \, , 
\label{eq:en2} 
\ee
where we have defined 
\be 
\beta \equiv {2 G \mob \over H_0^2 r_0^3} \, . 
\label{eq:beta} 
\ee
The parameter $\beta$ thus measures the effective ``strength'' of the 
galaxy or cluster. The requirement that equation (\ref{eq:en2}) have 
a turnaround point leads, as before, to a cubic constraint, which now 
takes the form 
\be
\vac \xi^3 - \beta \xi + (\om + \beta) = 0 \, , 
\ee
which in turn requires 
\be
\beta^3 \ge {27 \over 4} (\om + \beta)^2 \, \vac \, . 
\label{eq:bcubic} 
\ee
The minimum value of the strength parameter occurs when the equality
is saturated and we denote the corresponding value of $\beta$ by
$\bstar$.  Solving equation (\ref{eq:bcubic}), we find the value
$\bstar \approx 5.3$. We can thus determine the condition that must be
met in order for a test body to remain bound to an object of mass
$\mob$, i.e., 
\be
2 G \mob \ge \bstar H_0^2 r_0^3 \, . 
\label{eq:bound0} 
\ee
Inserting numerical values and scaling the result, we thus obtain 
the following criterion  
\be
{\mob \over 10^{12} M_\odot} \, > \, 3 \, h_{70}^2 \, 
\Bigl( {r_0 \over 1 \, {\rm Mpc}} \Bigr)^3 \, , 
\label{eq:bound} 
\ee 
where we have defined $h_{70} \equiv H_0$/(70 km s$^{-1}$ Mpc$^{-1}$).

As an immediate application of the above result
(eq. [\ref{eq:bound}]), we can determine whether or not the Milky Way
(Local Group) will remain bound to the Virgo cluster in our
cosmological future. The mass of Virgo is estimated to be $M_{\rm
Virgo}$ = $5 \times 10^{13} - 10^{14}$ $M_\odot$ and its current
distance from the Milky Way is about $r_0$ = 16 Mpc (e.g., Jacoby et
al. 1992). The requirement specified by equation (\ref{eq:bound}) is
not met --- the mass falls short by a factor of 100 --- so that the
Milky Way is not destined to be bound to Virgo. This result is
verified by numerical simulations (see \S 2 and Nagamine \& Loeb
2002).

Equation (\ref{eq:bound}) thus defines an effective sphere of
influence for any given astronomical object --- when considered as
isolated in a background universe dominated by a cosmological
constant. For our Milky Way galaxy, this sphere has radius $r_0
\approx 0.7$ Mpc.  For an isolated star (i.e., a free-floating star 
not associated with a galaxy), with typical stellar mass $M_\ast$ =
0.5 $M_\odot$, the sphere of influence has size $r_0 \approx$ 55 pc.
This size scale suggests that isolated binary star systems can remain
safely bound.  Although isolated pairs of galaxies can also remain
gravitational bound, they live much closer to the brink of instability. 

The above analysis finds the sphere of gravitational influence for
test bodies that are moving along with the Hubble flow at the present
epoch. However, due to the past evolution of the universe, bodies
that are now outside galaxies (or clusters) can be slowed down
relative to the Hubble flow due to the action of gravity in the past. 
In particular, the particles simulated in N-body simulations can be 
slowed down relative to the Hubble flow. To include this effect in 
the analysis, we modify equation (\ref{eq:en2}) to take the new form 
\be
\bigl( {d\xi \over d\tau} \bigr)^2 = A + (\om + \beta) 
\bigl( {1 \over \xi} - 1 \bigr) + \vac (\xi^2 - 1) \, , 
\label{eq:adef} 
\ee
where the constant $A$ represents the fact that the test particles 
have been slowed down relative to the Hubble flow (and rest of the 
quantities are the same as in eq. [\ref{eq:en2}]). The value 
$A=1$ corresponds to particles moving with the Hubble flow, so that 
particles that have been slowed will have $A < 1$. The case $A=0$ 
provides a benchmark, where test bodies have zero velocity (are 
turning around) at the present epoch. 

The requirement that equation (\ref{eq:adef}) have a turnaround 
solution implies a modified form of equation (\ref{eq:bcubic}), namely, 
\be
(\beta + 1 - A)^3 \ge {27 \over 4} \vac (\om + \beta)^2 \, . 
\label{eq:bcubicnew}
\ee
By solving the above cubic for a given $A$ when the inequality is 
saturated, we find the root $\beta^\star (A)$ that can be used to 
define the sphere of influence of a given cluster or galaxy 
according to equation (\ref{eq:bound0}). For the benchmark case, 
$A=0$, the root $\beta^\star\approx1.1$, and hence the effective sphere 
of influence is larger than the previous case by a factor of 1.7.  

In the following section, we compare these analytic predictions to the
results of numerical simulations. If we insert test bodies at the
present epoch --- at rest with respect to the Hubble flow --- then the
sphere of gravitational influence of existing structures is described
accurately by equation (\ref{eq:bound}). In addition, although mass
particles that begin at rest (with respect to the Hubble flow) in the
distant past are slowed down and have values $A < 1$ at the present
epoch, the prediction of equation (\ref{eq:bound}) works respectably
well (see the following section).

\section{NUMERICAL SIMULATIONS OF STRUCTURE FORMATION} 

To evaluate the analytic results of the previous section, particularly
equation (\ref{eq:bound}) defining the gravitational sphere for
influence for an object of mass $M$, we have run a series of numerical
simulations of the evolution of structure in a $\Lambda$-dominated
universe into the future.  All simulations were run using the GADGET
code (Springel, Yoshida, \& White
2001)\footnote{http://www.MPA-Garching.MPG.DE/gadget/} on a parallel
computer cluster at Michigan's Center for Advanced Computing.  A flat
\LCDM\ model is assumed, with matter density $\om = 0.3$,
power spectrum normalization $\sigma_8=1.0$, and $h = 0.7$, where the
Hubble parameter $H_0 = 100h
\kmsmpc$. The simulations followed the evolution of a 256.48
$h^{-1}{\rm Mpc}$ region with $128^3$ dark matter particles of mass
$6.70 \times 10^{11} h^{-1}M_\odot$, starting from $a=0.258$
($z=20.8$), through the present, and forward to $a = 100$.  A
gravitational force softening of $200 h^{-1}{\rm kpc}$ (in physical
units) was used throughout the computation.

In addition to the simulation described above, which was evolved to
$a$=100, we performed a similar simulation that evolved the $a=1$
configuration forward with an added set of initially stationary test
particles.  In this run, 96 particles were placed in a spherically
symmetric fashion around the centers of 19 dark matter halos. In each
case, they were placed at rest on one of 12 concentric spheres
centered at the most bound position of the selected group. This
simulation was also evolved to $a=100$.

\begin{figure*} 
\centerline{\epsfxsize=0.8\textwidth\epsffile{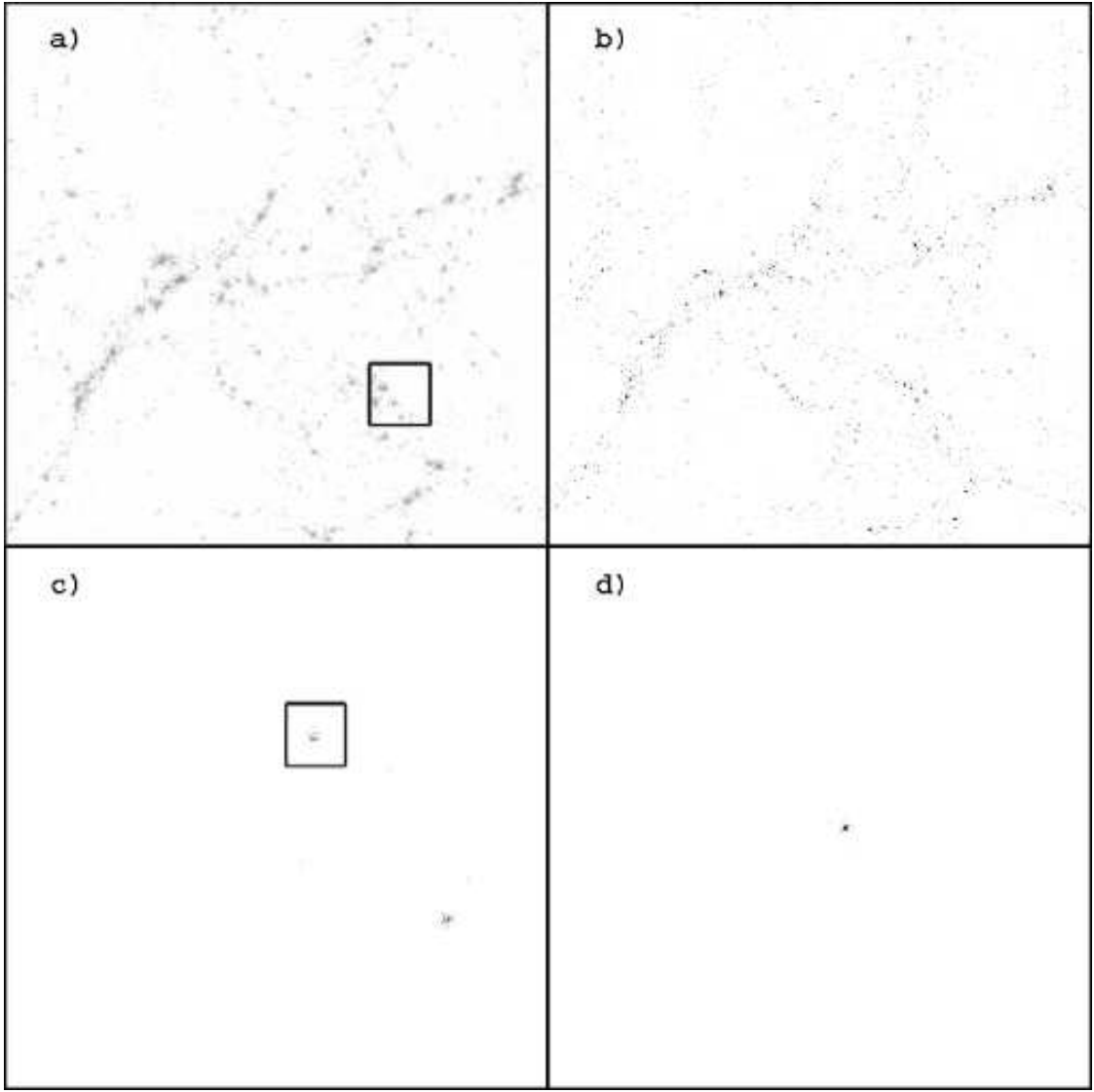}}
\caption{
Evolution of structure in a $\Lambda$-dominated universe. Panels a)
and b) show snapshots of a comoving region of the universe 128 $h^{-1}$
Mpc on a side and 25 $h^{-1}$ Mpc thick today ($a=1$) and at cosmic 
age 91.5 Gyr in the future ($a=100$), respectively.  Panels c) and d) 
show regions of the same {\it physical} size as that shown in panel a)
at epochs $a=11$ and $a=100$, respectively. The box in panel a)
locates the region shown in b), and the box in panel c) locates the
region shown in d).  
\label{fig:4slices} } 
\end{figure*}

The top two panels of Figure \ref{fig:4slices} show structure in a
fixed comoving region 128 $h^{-1}$ Mpc wide by 25 $h^{-1}$ Mpc deep at
(a) the present epoch and (b) $a=100$.  The large-scale pattern of the
cosmic web is well established by $a=1$ and evolves little thereafter
(as emphasized by Nagamine \& Loeb 2002).  Clusters of fixed physical
size shrink as $1/a$ in the comoving frame, so the halo population
effectively condenses into a sea of ``droplets'' embedded in the
frozen linear modes that define the filaments, walls, and voids 
that characterize the cosmic structure today. 

Panels (c) and (d) change perspective by showing how the physical
region of panel (a) appears at $a=11$ and $a=100$, respectively.  
The physical separation between bound structures grows exponentially
in time during the deSitter expansion phase of the dark energy dominated
era.  The future is increasingly lonely. 

To compute the size of a bound halo --- its gravitational sphere of
influence defined in the previous section --- we identify the smallest
radius at which the local mean radial velocity is significantly larger
than zero.  We measure such sizes using two different tracers: the
simulation particles themselves or the test particles of the second
simulation.  Since the latter are embedded at rest at $a=1$, their
future evolution should better follow, within the limits provided by a
monopole description of gravity, the prediction of equation
(\ref{eq:bound}) which assumes a pure, spherical Hubble flow.

\begin{figure*}
\begin{minipage}[b]{0.47\linewidth}
\centerline{\epsfxsize=\colwidth\epsffile{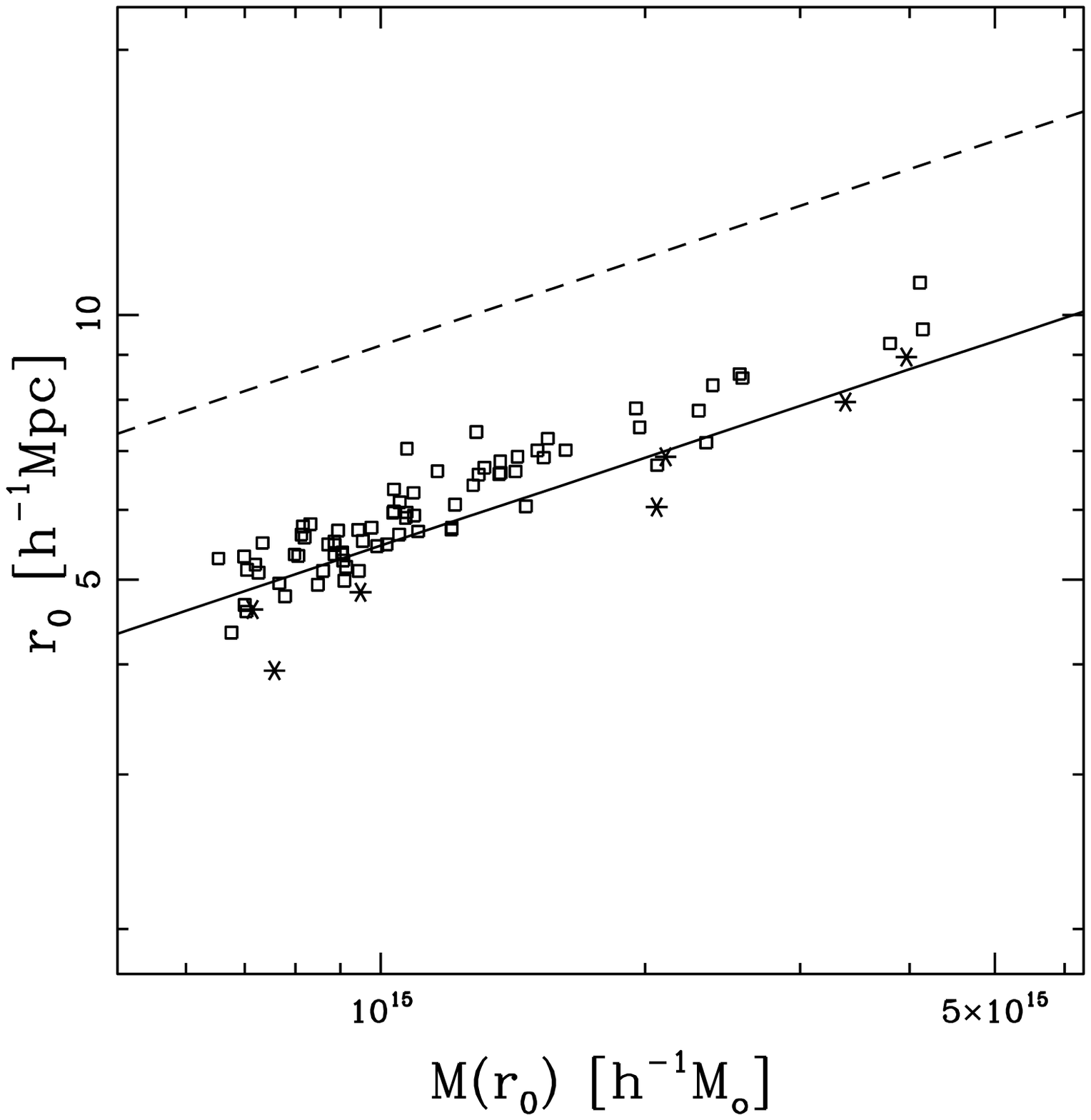}}
\caption
{Gravitational sphere of influence of a cosmological object, as a
function of the mass enclosed within the sphere of influence at a=1.
The lower solid line marks the analytic prediction for objects starting at
rest with respect to the Hubble flow (eq. [{\ref{eq:bound}}]); the upper
dashed line marks the analytic prediction for objects that are at rest
relative to the cluster at the present epoch (i.e., objects that are
marginally separated from the Hubble flow --- see
eq. [\ref{eq:adef}]).  These analytic predictions are compared with
simulations, both for cluster particles (open squares) and for the
test particles placed in the simulation at rest (stars).
\label{fig:soi}}
\end{minipage}
\hfill
\begin{minipage}[b]{0.47\linewidth}
\centerline{\epsfxsize=1.2\colwidth\epsffile{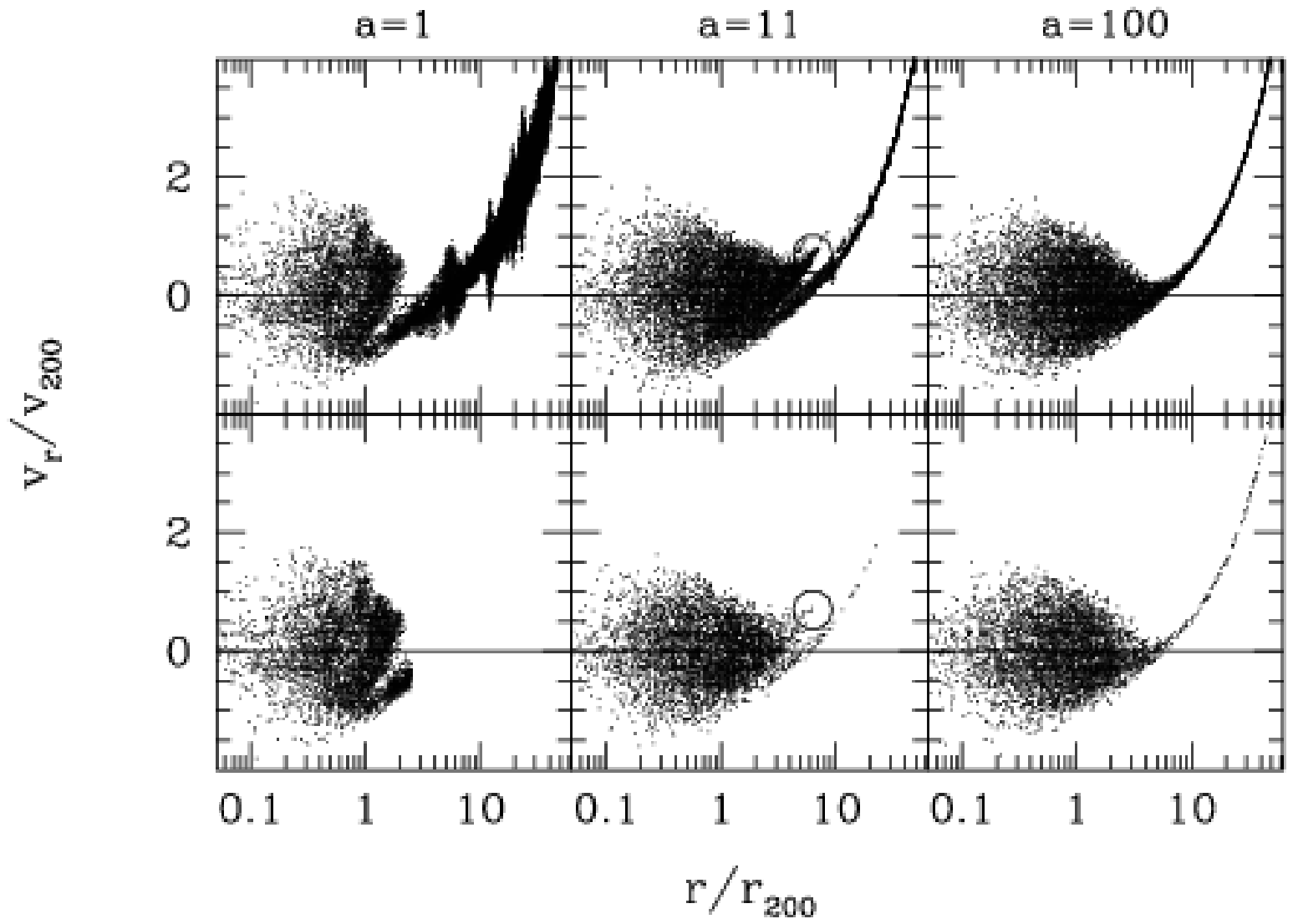}}
\vskip -2truecm
\caption{
Radial velocity relative to the cluster center as a function of 
distance for particles in the largest halo at the epochs indicated.  
The upper panel shows all matter while the lower panel shows only
those particles that lie within $2.5\rvir$ at $a=1$.
\label{fig:vradPart} 
}
\end{minipage}
\end{figure*} 

Measurements of the gravitational sphere of influence are shown in
Figure \ref{fig:soi}.  Note that the relevant mass is the mass within
the sphere of influence at the present epoch; this is typically a
factor of 2--3 larger than $\mvir$.  The relation between this mass
and its sphere of influence for test particles placed in the
simulation at rest at $a=0$, shown as stars, is in excellent agreement
with the predictions of equation~(\ref{eq:bound}), shown as the lower
line in the figure.  Values for the simulation particles, shown as
open squares, are slightly larger (by about $10\%$) because radial
infall in the weakly non-linear regime reduces the kinetic energy
associated with Hubble flow.  These sizes are bounded from above by
the modified relation, equation~(\ref{eq:adef}), with $A=0$, shown as
the upper dashed line in the figure.

\subsection{Halo Phase-space Structure} 

In the far future, the character of halos' radial phase-space
structure is markedly different from that of today.  The upper row of
Figure~\ref{fig:vradPart} shows the radial velocity pattern of the
most massive halo at $a=1$, 11 and 100.  The halo identified at
present remains the most massive at all future times.  At each epoch,
lengths are expressed in units of \rvir, the radius within which the
mean enclosed density is 200 times the critical value.  Physical
velocities are expressed in units of $\vvir=\sqrt{G\mvir/\rvir}$,
where \mvir\ is the mass within \rvir.  The halo's physical size grows
from $\mvir= 1.08 \times 10^{15} \hinv\msol$ at $a=1$ to $2.35\times
10^{15} \hinv\msol$ at $a=11$, and remains nearly constant thereafter.

At $a=1$, positive velocity particles at radii $r/\rvir \sim 1-2$
represents material that has penetrated the halo core and is streaming 
outward to large radii.  This ``processed'' material mixes at these
radii with a fresh stream infalling at $v_r \simeq -\vvir$. The stream 
extends outward through a zero-velocity surface at $r \simeq 5\rvir$,
and asymptotically approaches the Hubble flow at large radii.  The
present-epoch interior phase structure displays a lumpy morphology,
evidence of recent merger activity that has not yet dynamically
relaxed.  The infall pattern shows spikes of enhanced dispersion, the
signatures of neighboring halos. Those within the gravitational sphere
of influence are destined to merge with the central halo.

By $a=11$, phase mixing has smoothed the interior structure
considerably, but the remains of a recent merger can be seen in a
small tail of outflow centered at $r/\rvir \sim 6$ (circled in the
panel).  The infall regime is much quieter, with only a few collapsed
structures evident in the tail of Hubble flow.  At $a=100$, the
interior phase structure is extremely homogeneous and the infall
regime is essentially silent; no collapsed structures perturb the flow
within 50\rvir.

The lower panels of Figure~\ref{fig:vradPart} show how the non-linear
material of the present cluster evolves into the future. Only those
particles lying within $2.5\rvir$ at $a=1$ are shown.  This radius is
chosen to encompass all the ``processed'' halo material at the present
epoch.  Although this boundary extends beyond most common choices for
the ``virial radius'' of a halo (Evrard, Metzler \& Navarro 1996; Eke,
Cole \& Frenk 1996; White 2001), it is evident that some material
currently within this radius is destined for future escape.  A thin
tail forms of material lifted off the halo, representing $0.8$ and
$2.6$ percent (at $a=11$ and 100, respectively) of the mass within
$2.5\rvir$ at $a=1$.  Note that the merger system at $a=11$ circled in
the upper panel is not as strongly evident in the lower panel,
indicating that it lies beyond $2.5\rvir$ at the present epoch.

The simple, smooth phase structure at late times suggests a long-term
equilibrium.  In the next subsection, we address the question of the
eventual shape of the radial mass profiles of halos.

\subsection{Halo Density Profile} 

Numerical studies have revealed that the non-linear density structure of
halos formed through gravitational clustering takes on a common form. The 
first such studies (Navarro, Frenk \& White 1996; 1997) showed that this 
density profile can be written in the form 
\be
\rho(r) = {4\rho_s \over r/r_s [1 + (r/r_s)]^2 } \, , 
\label{eq:rhoNFW} 
\ee
where $r_s$ is a characteristic radius where the logarithmic slope of
the density profile is isothermal: ${\rm d ln}\rho/{\rm d ln} r = -2$.
The ratio $\rvir/r_s$ defines the concentration parameter $c$.  Note
we have chosen the convention $\rho_s = \rho(r_s)$; in this case the
inner density can be directly related to the concentration of an NFW
profile via
\be
\rho_s = \frac{1}{4}\delta_c\rho_{\rm c}
 = \frac{\Delta}{12}\frac{c^3}{\rm{ln}(1+c)-c/(1+c)}\rho_{\rm c}.
\ee
Previous fits to this
density profile have been limited to radii $r \lsim \rvir$ and epochs
$a \leq 1$.  We present here an extension of this form to larger radii
and future epochs.

Figure~\ref{fig:profile} shows radial density profiles derived from
stacking the 50 most massive halos at each epoch displayed.  Solid
lines in the figure show binned profiles for five epochs from
$a=1--100$, while dashed lines show fits to the form
\be
\rho(r) = {A\rho_s \over r/r_s [1 + (r/r_s)^p]^{3/2} }
[1 + r/r_\infty]^{1 + 3p/2} \, .
\label{eq:rhopro} 
\ee
Here $r_s$ is a scale radius similar to the that of
equation (\ref{eq:rhoNFW}), $r_\infty$ is an asymptotic radius, $p$ is
a free parameter and $A=2^{3p/2}/[1 + r_s/r_\infty]^{1+3p/2}$.  We find
that values $p=1.8$, $r_s = 0.50$ and $r_\infty = 4.7\rvir
a^{6/(3p+2)}$ provide fits that are accurate to $\langle
(\delta\rho/\rho)^2 \rangle^{1/2} \sim 35\%$ over the full range of
radii and epochs examined.  The scaling of $r_\infty$ with expansion
factor ensures that the profile approaches the mean mass density of
the universe as $r \rightarrow \infty$.

The profile of equation (\ref{eq:rhopro}) is steeper than the NFW form
at radii beyond $\rvir$.  As $r_\infty \rightarrow \infty$, the
logarithmic slope of the profile well beyond \rvir\ approaches $3.7$
(shown by the dot-dashed curve in Figure \ref{fig:profile}), steeper
than the slope of 3 from the NFW case. This difference in slope keeps 
the enclosed mass from being logarithmic divergent, as implied by 
a formal extrapolation of the NFW form (compare eq. [\ref{eq:rhoNFW}]
with [\ref{eq:rhopro}]). 

In fact, the mass of a halo in the far future will be simpler to
define than it is today.  At present, radial infall and incomplete
dynamical relaxation make the choice of the ``edge'' of a cluster
somewhat arbitrary (White 2001; Evrard \& Gioia 2002).  In the
relatively near future, however, halos evolve toward an equilibrium
configuration that is bounded by an increasingly sharp zero-velocity
surface (Figure~\ref{fig:vradPart}).  Ultimately, a meaningful and
unique definition of mass emerges, namely, all of the matter lying 
interior to this well-defined zero-velocity surface.  In addition, as shown by 
Figure \ref{fig:profile}, the density profile attains a well-defined form.

\begin{inlinefigure}
\centerline{\epsfxsize=\colwidth\epsffile{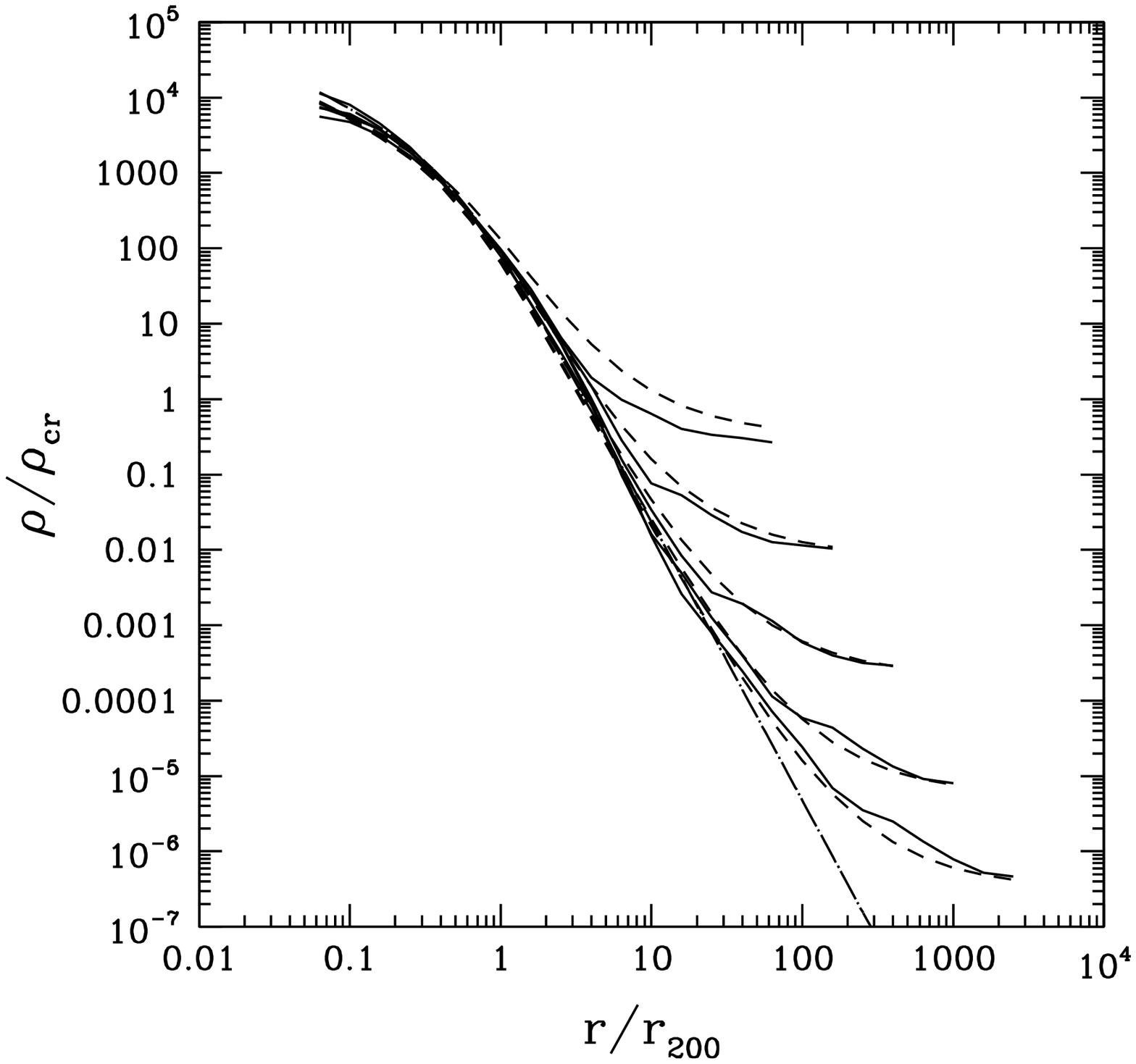}}
\figcaption
{The asymptotic form for the density distribution of dark matter
halos. The solid curves show the (nearly) universal form for dark
matter halos, for a variety of epochs (from top to bottom, $a$ = 1,
3.38, 11.4, 38.6, and 100). Starting just after the present time, the
dark matter halo profiles show essentially the same form, only the
outer boundary is stretched to match onto the ever-lower density of
the background universe. The dashed curves show the fit to the
numerical results, equation (\ref{eq:rhopro}), evaluated at each epoch
with the parameters given in the text. The dot-dashed curve shows the
asymptotic form of the density profile in the limit $t \to \infty$. 
\label{fig:profile}
}
\end{inlinefigure}

\section{LONG-TERM RAMIFICATIONS OF COSMIC ACCELERATION} 

The considerations of the previous sections outline the requirements
necessary for future structure formation and the criteria for test
bodies to remain bound to existing astronomical structures. With these
results in place, we can fill out the picture of the future evolution
of our cosmos. In particular, we can determine the time scales for
various bound structures to grow isolated, the corresponding effects
on the background radiation fields in the universe, and the freezing
out of particle annihilation processes.

\subsection{Isolation of bound structures} 

Because our universe contains a dark vacuum component, it has a
well-defined horizon scale. For unbound objects living in this
accelerating universe, the next issue is thus to determine when the
objects leave the horizon. The background universe can be described
by its line element, which can be written in advanced time coordinates 
in the form 
\be 
ds^2 = - \bigl[ 1 - \chi^2 r^2 \bigl] dv^2 + 2 dv dr + r^2 d \Omega^2 \, . 
\ee
This line element ignores the curvature due to gravitationally-condensed 
structures (see the following subsection). 
The parameter $\chi$ is related to the magnitude of the cosmological 
constant and is defined by the relation (in natural units) 
\be 
\chi = \Bigl({2 \pi^3 \over 45}\Bigr)^{1/2} {\cct^2 \over \mpl} \, , 
\ee 
where $\cct$ is the effective temperature scale of the 
cosmological vacuum energy ($\cct$ $\approx$ 0.003 eV $\approx$ 
35 K for the presently suspected cosmological constant). 
In such a universe, the horizon distance $r_H$ is given by 
\be
r_H = \chi^{-1} = {c \over H_0} {2 \over \pi} 
\Bigl( {15 \over \vac} \Bigr)^{1/2} 
\approx 12,600 \, \, {\rm Mpc} \, , 
\label{eq:horizon} 
\ee
where the second equality assumes the standard values $\om$ = 0.3 and
$\vac$ = 0.7. This horizon distance $r_H$ is not the same as the
particle horizon, but rather is essentially the Hubble radius. This
distance scale $r_H$ provides an effective ``boundary for microphysics'' 
within the much larger space-time of the universe (for further discussion
of horizons, see Kolb \& Turner 1990 and Ellis \& Rothman 1993).

For a flat universe with both matter and vacuum components, the 
scale factor $a(t)$ increases according to the function 
\be
a(t) = \Bigl( {\om \over 1 - \om} \Bigr)^{1/3} 
\Bigl\{ \sinh \bigl[ {3\over2} \sqrt{1 - \om} \, H_0 t \bigr] 
\Bigr\}^{2/3} \, . 
\label{eq:scale}
\ee
At later times, the scale factor approaches the simpler 
asymptotic form 
\be
a_{\rm asym} (t) = \bigl( {\om \over 4 \vac } \bigr)^{1/3} 
\exp [ \sqrt{ \vac } \, H_0 t ] \, . 
\label{eq:asymp}
\ee 
One can see immediately that the e-folding time scale of the future
universe is $\tau_e$ = $\vac^{-1/2} H_0^{-1}$ $\approx$ 17 Gyr.
Furthermore, the scale factor $a(t)$ approaches this asymptotic form
on an even shorter time scale. If the full expression 
(eq. [\ref{eq:scale}]) is written as the asymptotic form
(eq. [\ref{eq:asymp}]) plus correction terms, those correction terms
decay with a time scale $\tau = (3 H_0 \sqrt{\vac})^{-1}$ = $\tau_e/3$
$\approx$ 5.6 Gyr. This time scale --- somewhat longer than the age of
the solar system and appreciably younger than the current age of the
universe --- is a direct manifestation of the cosmological constant
problem.

Given that any extant cosmic structure has a sphere of influence
(eq. [\ref{eq:bound}]) and that the universe has a fixed horizon size
$r_H$, every structure will become isolated when the radius of its
sphere of influence is stretched beyond the horizon, i.e., when $a(t)
r_0 > r_H$. For a structure of mass $\mob$, isolation occurs at a time
$t_{\rm iso}$ given by 
\be
t_{\rm iso} = {2 \over 3} \tau_e \, 
\sinh^{-1} \Biggl\{ \Bigl[ {4 \bstar (15)^{3/2} \over \pi^3} 
{c^3 \tau_e \over G \mob \om} \Bigr]^{1/2} \Biggr\} \, . 
\label{eq:isolate} 
\ee 

As a result, for vast majority of cosmological time, the cosmos will
be divided into ``island universes'' in the sense that bound clusters
of galaxies will retain their sizes (a few to several Mpc) while the
distance between clusters grows exponentially (see also Chieuh \& He
2002). Given typical cluster sizes and separations, we predict that
clusters will grow isolated in about 120 Gyr. For the particular
values appropriate for the nearby Virgo cluster, equation
(\ref{eq:isolate}) implies an isolation time of 132 Gyr (this time is
the age of the universe at the time of isolation; since the universe
is already about 14 Gyr old, this event will occur 118 Gyr from now).
Structures with lower mass have smaller spheres of gravitational
influence and require longer times to grow isolated. Our local group,
with an estimated mass of $M_{LG} \approx (2.3 \pm 0.6) \times
10^{12}$ $M_\odot$ (van den Bergh 1999), will be isolated at cosmic
age $t$ = 175 Gyr. As another example, a star that is not
gravitationally bound to a larger structure (e.g., one that has been
scattered out of a galaxy) requires somewhat longer to become isolated
--- about 336 Gyr. For comparison, the lifetimes of the smallest,
longest-lived stars are measured in tens of trillions of years
(Laughlin, Bodenheimer, \& Adams 1997; hereafter LBA97), about one
hundred times longer than the isolation time for galaxy clusters. For
most of eternity, and indeed for most of the Stelliferous Era,
clusters will be alone. Inside the galaxies, the expansion has
essentially no effect, and star formation and stellar evolution
continue for trillions of years (AL97). When viewed on the large
scale, however, these clusters will behave like point sources, pumping
radiation into an ever-expanding void.

\subsection{Asymptotic structure of space-time}

In the long term, existing cosmic structures will remain bound,
but will grow isolated. These structures will be embedded within an
accelerating universe with a constant horizon scale (eq.
\ref{eq:horizon}). This process effectively divides the present-day
universe into many smaller regions of space-time. These ``island
universes'' display properties of a universal nature.

Every given ``island universe'' will approach a fixed overdensity.
The energy density contained with the horizon scale is equivalent to a
mass $M_H$ given by
\begin{eqnarray}
\lefteqn{M_H = {4 \pi \over 3} \rho_{V (t \to \infty)} \, r_H^3 =}
\nonumber \\
&&  
{\vac \over 2 G H_0} \, \Bigl( {2 c \over \pi} \Bigr)^3 \,
\Bigl( {15 \over \vac} \Bigr)^{3/2} \approx
8 \times 10^{23} M_\odot \, .
\label{eq:hormass} 
\end{eqnarray}
Since the mass contained within any given isolated cluster will be
constant, the overdensity approaches a constant value.  For our
particular environment, the local group will remain bound with its
mass of about $M_{LG}$ = 2.3 $\times 10^{12}$ $M_\odot$. The mass in
our local region is thus destined to be a minor perturbation on the
cosmos itself, even within our local island universe. For cosmic ages
older than 175 Gyr, the mass contribution to the universe contained
within the local group is given by $\delta_\infty = M_{LG} / M_H
\approx 3 \times 10^{-12}$, only 3 parts per trillion.

Our numerical simulations indicate that cosmic structures, from
galaxies to clusters, tend to develop universal forms for their
density profiles.  As a result, every island universe will attain the
same general form for its space-time. In particular, since the density
profile attains the universal form described by equation (\ref{eq:rhopro}), 
the line element $ds^2$ for the space-time within the horizon distance
$r_H$ also attains a universal form. If we take the center of the
coordinate system to be the center of the cluster and assume that the
mass distribution is spherically symmetric, the line element can be
written in the form 
\begin{eqnarray}
\lefteqn{ds^2 = - \Bigl(1 - A(r) - \chi^2 r^2 \Bigr) dt^2}
\nonumber \\
&& + 
\Bigl(1 - B(r) - \chi^2 r^2 \Bigr)^{-1} dr^2 + r^2 d\Omega^2 \, ,
\label{eq:linein} 
\end{eqnarray}
where $A(r)$ and $B(r)$ depend on the mass distribution (see, e.g.,
Misner, Thorne, \& Wheeler 1973).  This form for the line element is
that of a mass distribution embedded in deSitter space (see also
Bardeen 1981, Mallet 1985, Chiueh \& He 2002). As a result, there
exists an outer horizon at $r = \chi^{-1}$.  We provide a more detailed
specification of the metric in a future work (Adams et al 2003).

The outer horizon supports the emission of radiation through a
Hawking-like mechanism (e.g., Fulling 1977, Birrell \& Davies 1982).
As a result, the universe will be filled with a nearly thermal bath of
radiation with characteristic wavelength $\lambda \sim r_H \sim
\chi^{-1} \sim 12,600$ Mpc and characteristic temperature $T \sim \chi
\sim \Lambda^2/\mpl \sim 10^{-33}$ eV $\sim 10^{-29}$ K. This bath of
radiation will become the dominant background radiation field at very
late times (after about one trillion years -- see the following
section).

\subsection{Background radiation fields in an accelerating universe} 

As the universe expands, all radiation fields are redshifted to longer
wavelengths. An important milestone is reached when the typical
wavelength of a given radiation field grows longer than the
cosmological horizon scale defined by equation (\ref{eq:horizon}). 
After this crossing, the photons are larger than the largest ``box'' 
that the universe has to contain them. For later times, it no longer 
makes sense to describe the photons in terms of a distribution function. 
Inside the horizon, in the limit $\lambda \gg r_H$, the background 
photons will appear as ''slowly'' varying electric fields rather than as 
particles of light. The dominant background radiation field will be 
that produced by the horizon itself through a Hawking-like mechanism 
(see \S 4.2, Fulling 1977, Birrell \& Davies 1982). 

Given the scale factor of the universe and the present day wavelength 
of a radiation field, it is straightforward to find the time at which 
the photons are stretched beyond the horizon scale, i.e., when 
$\lambda a(t) > r_H$. For the cosmic background radiation, the 
present day wavelength (at the peak of the distribution) is about 
$\lambda_0$ = 0.1 cm and the photons cross the horizon at a time of 
1120 Gyr. 

The cosmic background photons are stretched beyond the horizon well
before the stars stop shining. Star formation and stellar evolution
will continue until the universe is tens of trillions of years old
(AL97, LBA97). Suppose that stars continue to shine for 10 trillion
years. By this late epoch, most of the remaining stars will be red
dwarfs that emit light with a characteristic wavelength of $\lambda$ =
1 $\mu$m = $10^{-4}$ cm.  If this red light is emitted up to a time
$t_\ast$ ($\approx$ $10^{12}$ yr) and observed at a later time $t$,
its observed wavelength is given by 
\be
\lambda_{\rm obs} = \lambda_{\rm emit} 
{ a_{\rm obs} \over a_{\rm emit} } = \lambda_{\rm emit} 
\exp\big[ (t-t_\ast)/\tau_e \bigl] \, , 
\label{eq:stretch} 
\ee
where $\tau_e$ = 17 Gyr is the e-folding time of the future universe.
Starlight leaves the horizon when $\lambda_{\rm obs} > r_H$. Using
this criterion in conjunction with equations (\ref{eq:horizon}) and
(\ref{eq:scale}), we find that starlight is redshifted out of the
horizon over a time interval of only $\Delta t$ = $(t-t_\ast)$ = 1260
Gyr. Stellar evolution times --- for the smallest stars --- are much
longer than the cosmological expansion times, so that photons are
rapidly stretched beyond the horizon. Specifically, this stretching
time is a small fraction of the Stelliferous Era, the time over which
the universe will contain substantial numbers of hydrogen burning
stars, i.e., $\Delta t/t_\ast \approx 10^{-2}$.
 
\subsection{Particle annihilation in an accelerating universe} 

For material between galaxies --- particles that are not bound to 
large structures --- future evolution can continue through particle 
annihilation. The number density $n$ of a given particle species 
is given by the evolution equation 
\be
{dn \over dt} + 3 H n = - \sig n^2 \, , 
\label{eq:anni} 
\ee 
where $\sig$ is the appropriate average of the interaction cross
section and the relative velocity (see, e.g., Kolb \& Turner 1990; see
also Cirkovic \& Samurovic 2001).  For an accelerating universe with a
cosmological constant, the solution to equation (\ref{eq:anni}) can be
found and written in terms of the scale factor, i.e., 
\be
n(a) =  n_0 a^{-3} \Biggl\{ 1 + {2 \Gamma \over 3 \om} 
\Bigl[ 1 - \sqrt{\vac + \om a^{-3}} \Bigr] \Biggr\}^{-1} \, , 
\ee
where $n_0$ is the particle density at the present epoch and 
where we have defined $\Gamma \equiv \sig n_0 / H_0$. The leading 
factor ($a^{-3}$) represents the dilution of the number density 
due to cosmic expansion, whereas the second factor in brackets 
incorporates the effects of continued particle annihilation. In 
an accelerating universe, annihilation is highly suppressed. 
For example, in the asymptotic limit $t \to \infty$, $a \to \infty$, 
this factor becomes $\cal F$ = 1 + $2 \Gamma (1-\sqrt{\vac})/3\om$
$\approx$ 1 + 0.0048 [$\sig$/barn$\cdot c$]. For electron-positron 
annihilation, for example, the maximum correction term is less than a
percent. For annihilation of cold dark matter particles (thought to be
the dominant matter contribution), the interaction cross sections
typically lie in the range $\sigma \sim$ $10^{-12} - 10^{-14}$ barn
(e.g., Kolb \& Turner 1990) and the speed $v/c \sim 10^{-3}$. As a 
result, the already small correction term (0.0048) is suppressed by an
additional 16 orders of magnitude. This enormous suppression is driven 
by the relentless expansion of an accelerating universe. 

When the number density grows so diffuse that the universe contains
less than one particle per horizon volume, then individual particles
are effectively isolated. The condition for such isolation can be
written in the form 
\be
a(t) > \Bigl( {4 \pi \over 3} n_0 \Bigr)^{1/3} \, r_H \, , 
\ee
where the scale factor $a$ is given by equation (\ref{eq:scale}) 
and the horizon scale is given by equation (\ref{eq:horizon}).
Adopting a present day number density of $n_0$ = $10^{-6}$ 
cm$^{-3}$ results in a particle isolation time scale of about 1060 Gyr.

\section{DISCUSSION AND CONCLUSIONS} 

This paper explores the future evolution of a universe dominated by
dark vacuum energy.  Analytic estimates are compared to results of
numerical simulations that follow the evolution of future structures
in such an accelerating universe.

For a universe with cosmological constant $\vac=0.7$, only those
regions with present-day overdensities $\delta_0 > 17.6$ will remain
gravitationally bound, in agreement with earlier estimates (Lokas \&
Hoffman 2002; Nagamine \& Loeb 2003).  We generalize this result to
include quintessence models with constant forms for the equation of
state (Appendix A). We have also derived the condition required for
test bodies to remain bound to existing structures (see eq.
[\ref{eq:bound}]) and verified its validity with numerical simulations 
(to within $\sim 10\%$; see Figure \ref{fig:soi}).  Any collapsed object --- from a star
to galaxy cluster --- has a finite sphere of gravitational influence
in an accelerating universe, with radius $r_0 \approx$ 1 Mpc $(M_{\rm
obj}/10^{12} M_\odot)^{1/3}$. For quintessence models, we have derived
an analogous result for the sphere of gravitational influence (see
Appendix A and eq.  [\ref{eq:fit3}]).

From a co-moving perspective, the large-scale appearance of the future
universe is little changed from that of today.  Matter in the cosmic
web drains efficiently into collapsed halos that shrink in comoving
coordinates.  The halos are essentially frozen in place while their
contrast relative to the mean background grows with time. In physical
coordinates, the view is rather different.  The vast majority of the
galaxies now visible are pulled out of the immediate horizon of any
given bound structure (a cluster or group).  In the long term, only
the cluster or group itself remains within the effective horizon scale
of $r_H$ = 12,600 Mpc.

The long-term structure of space-time consists of a flat metric
dimpled with isolated clusters that approach a fixed mass profile.  
We find that halo density profiles approach a form similar to, but
steeper at large radii than, the NFW profile (equation
[\ref{eq:rhopro}]). It is important to emphasize that {\sl every\/}
halo grows isolated in the long term, i.e., every gravitationally
bound mass concentration ultimately becomes the only structure within
its own island universe.  In each such local region, the halo density
takes the form shown in Figure \ref{fig:profile} and the line element
of the space-time metric takes the form given by equation
(\ref{eq:linein}).  Although the halo mass varies from region to
region, the form of the metric -- and hence the geometry of space-time
-- is nearly universal. In all cases, the halo mass in any region
provides only a minor contribution to the overall mass/energy budget,
with $\mob/M_H \sim 10^{-11}$.

\begin{table*}
\caption{Time Scales and Scale Factors} 
\begin{center}
\begin{tabular}{lcc}
\tableline
\tableline
{\sl Event} & {\sl Time} $\tau$ (Gyr) & $a(\tau)$\\ 
\tableline 
Time scale for scale factor to approach exponential form & 5.6 & -- \\ 
Inverse Hubble constant $H_0^{-1}$ & 14 & -- \\
e-folding time of the future universe $(H_0 \sqrt{\vac})^{-1}$ & 17 & -- \\
\tableline 
Current age of the universe & 13.7 & 1 \\ 
Virgo Cluster leaves our horizon & 132 & 1000\\ 
The Local Group grows isolated & 180 & $2 \times 10^4$\\ 
Exiled stars become isolated & 336 & $2 \times 10^8$\\ 
Individual particles grow isolated & 1060 & $6 \times 10^{26}$\\ 
CBR photons stretch beyond the horizon & 1120 & $2 \times 10^{28}$\\
Optical photons stretch beyond the horizon & 1260 & $10^{32}$\\
Lifetime of longest-lived stars & 17,000 & $10^{434}$ \\
End of the Stelliferous Era & 100,000 & $10^{2554}$\\
\tableline 
\label{table}
\end{tabular}
\end{center} 
\end{table*}

As the universe continues to expand, and accelerate, cosmic radiation
fields are redshifted to increasingly long wavelengths. After about
one trillion years, the cosmic background radiation (leftover from the
big bang) is stretched beyond the horizon and the dominant radiation
background is that emitted by the horizon itself through a 
Hawking-like mechanism.  Many of the results of this investigation can
be summarized in terms of the relevant time scales, which are listed
in Table \ref{table}. To emphasize the mismatch between the various
time scales, the table also lists the scale factor for each relevant
epoch. For example, individual stars grow isolated in 336 Gyr ($a$ = 
$2 \times 10^8$), but the longest-lived stars burn hydrogen for 
17,000 Gyr (when $a$ = $10^{434}$). 

With the analysis complete, many of the time scales and length scales
of the future universe can be understood in simpler terms. A
dimensional analysis -- presented in Appendix B -- shows that the most
important time scale is given by the asymptotic form for the Hubble
parameter $H_\infty$ = $\sqrt{\vac} H_0$ $\approx$ 59 km s$^{-1}$
Mpc$^{-1}$. The accelerated expansion itself completely dominates the
evolution of the universe as a whole, so that all of the time scales
are determined by $H_\infty^{-1}$ $\approx$ 17 Gyr and logarithmic
multiplying factors (see Appendix B). By comparison, time scales for
stellar evolution ($10^{13} - 10^{14}$ yr; AL97, LBA97) and dynamical
relaxation of galaxies ($10^{20}$ yr; BT87) are much longer. \\

The authors would like to thank Greg Laughlin for useful
discussions. This work was supported by the University of Michigan
through a Regents Fellowship (MTB), a Physics Department Fellowship
(RHW), by the Michigan Center for Theoretical Physics, and by NASA
Astrophysics Theory Grant NAG5-8458.

\appendix
\section{QUINTESSENCE} 
\bigskip

In this Appendix we generalize our results to the case of a vacuum
energy that depends on time, or equivalently, the scale factor $a$.
For the sake of definiteness, we adopt the standard form for the
vacuum energy equation of state, i.e., the vacuum pressure is given by
\be
p_{\rm vac} = w \rho_{\rm vac} \, , 
\ee  
where the parameter $w$ is constant and lies in the range $-1 \le w < 0$. 
Current observations seem to indicate a somewhat smaller range $-1 < w
\lsim -0.5$ (e.g., Limin et al. 2000, Balbi et al. 2001). In fact, 
Spergel et al (2003) place a 95\% confidence limit of $w \le -0.5$
using a combination of the WMAP CMB data and the HST key project
(Freedman et al. 2001) value for the Hubble constant and find that 
$w \le -0.78$ when additional constraints are added (from the
SNIa-derived redshift distance relation, the 2dFGRS large-scale
structure, and Lyman-$\alpha$ data, assuming a flat universe with
constant $w$). For completeness, we consider here the full range of
constant $w$ values from 0 to $-1$.  With this equation of state, the
scale factor evolves according to 
\be 
\Bigl( { {\dot a} \over a} \Bigr)^2 = H_0^2
\Bigl\{ \om a^{-3} + \vac a^{-p} \Bigr\} \, 
\ee
where the index $p = 3(1+w)$. 

The energy equation (\ref{eq:en2}) that determines whether or not
overdense regions collapse, and the fate of test bodies, can be
written in the form 
\be
\bigl( {d\xi \over d\tau} \bigr)^2 = 
\vac a^{-p} \xi^2 - \beta + (\om + \beta)/\xi \, , 
\label{eq:qenergy}
\ee
where $\beta$ measures the gravitational influence of an existing
structure according to equation (\ref{eq:beta}).  The case of
overdense regions can be considered by replacing $\beta$ with 
$\om \delta_0$, where $\delta_0$ is the overdensity. 

\begin{figure*}[t]
\begin{minipage}[b]{0.47\linewidth}
\centerline{\epsfxsize=\colwidth\epsffile{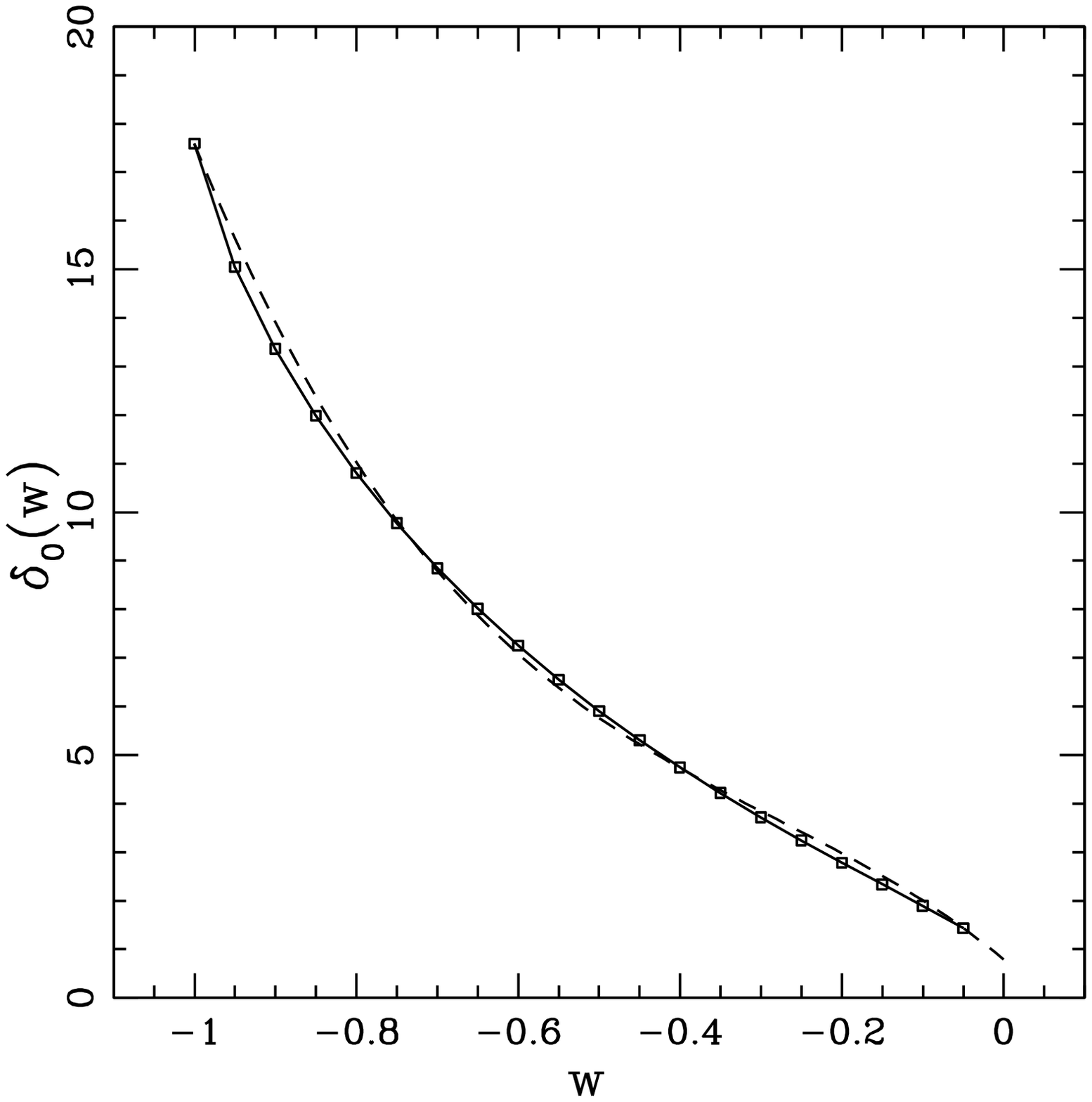}} 
\caption
{This plot shows the overdensity required for the collapse of future
structures as a function of the parameter $w$ appearing in the
equation of state for quintessence models. The sphere of gravitational
influence for existing cosmological structures is given by $r_G$ = $(2
G \mob / \beta^\star H_0^2)^{1/3}$, where the parameter $\beta^\star$
is related to the overdensity required for collapse of existing
regions via $\beta^\star = \delta_0 \om$ (see text).  The solid curve
shows the numerically determined values; the dashed curve shows a
cubic fit to the function (using eqs. [\ref{eq:fit1} -- \ref{eq:fit3}]). 
As $w$ becomes more negative, the overdensity required for collapse
becomes larger and the sphere of gravitational influence (for existing
structures) grows smaller --- the formation of future structure is 
more suppressed for smaller $w$. 
\label{fig:soiw}}
\end{minipage}
\hfill
\begin{minipage}[b]{0.47\linewidth}
\centerline{\epsfxsize=\colwidth\epsffile{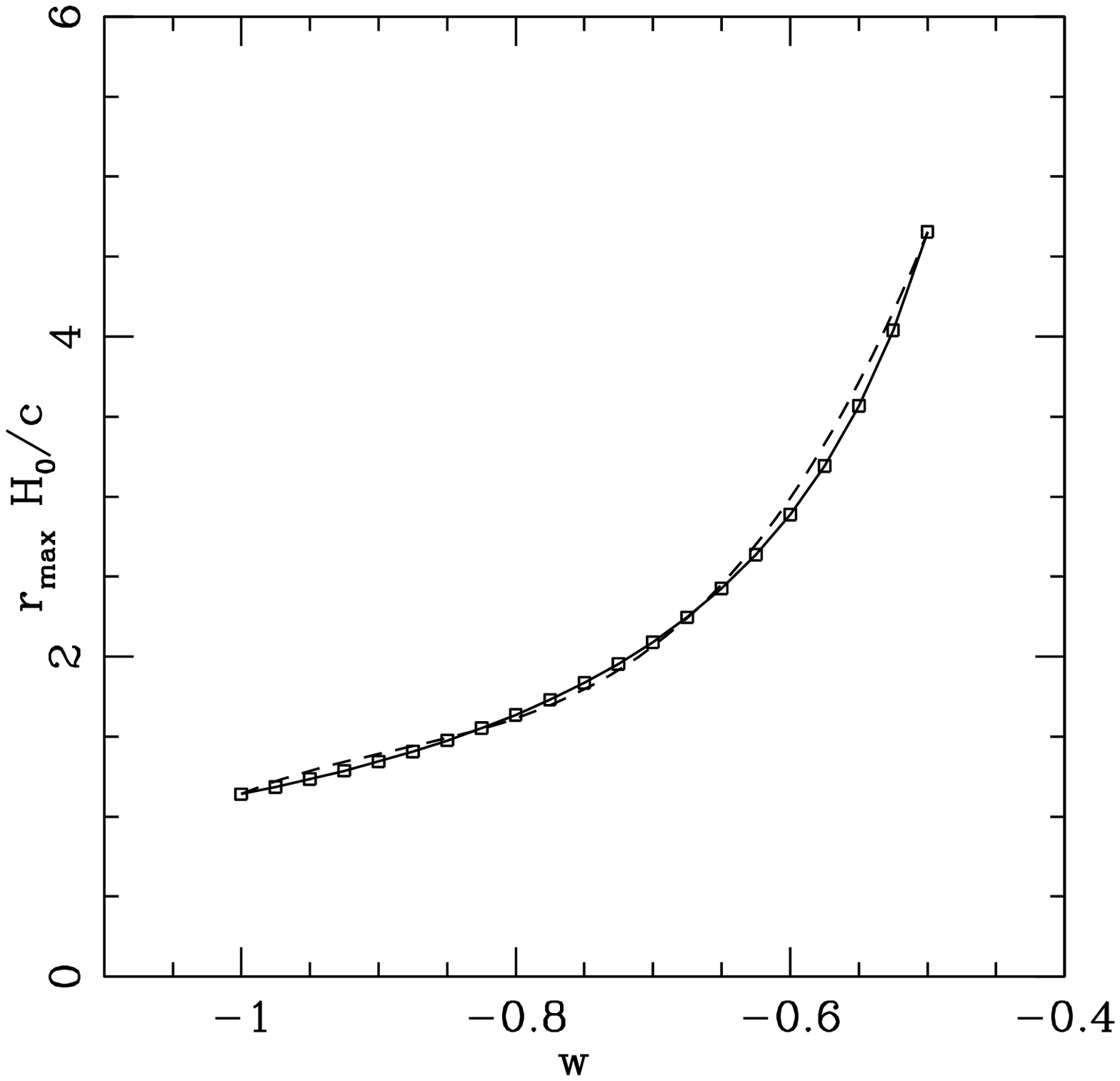}} 
\caption
{
The maximum distance that a light signal can propagate
between the present epoch and temporal infinity for an
accelerating universe described by equation of state parameter
$w$. The dashed curve shows a cubic fit to the numerically obtained
result (see Appendix A and eq. [\ref{eq:rmax}]).
\label{fig:rmax}}
\end{minipage}
\end{figure*}

In order to determine whether trajectories turn around (and hence
remain bound or collapse), the right hand side of equation
(\ref{eq:qenergy}) must vanish as before. In this case, however, the
resulting cubic equation has time dependent coefficients and so the
evolution of the scale factor must be considered simultaneously. As a
result, we combine the two equations by changing the independent
variable to $a$ and thereby obtain 
\be
\bigl( {d\xi \over da} \bigr)^2 = { 
\vac a^{-p} \xi^2 - \beta + (\om + \beta)/\xi \over 
a^2 \bigl[ \om a^{-3} + \vac a^{-p} \bigr] } \, . 
\label{eq:total} 
\ee
Trajectories turn around when the right hand side of this equation
vanishes. The minimum value of $\beta$ required for such turnaround
occurs when the right hand side of the equation has a double zero
(both the right hand side and its derivative with respect to $\xi$
vanish). To find the critical value of $\beta$, denoted here as
$\beta^\star$, we numerically integrate equation (\ref{eq:total}) and
iterate to find the value that provides a double zero. This procedure
must be carried out for every value of $w$ (or $p$). The resulting
values of $\beta$ are given below. The overdensities required for 
the collapse of future structures, for a cosmology with a given 
value of $w$, are given by $\delta_0 = \beta^\star / \om$. This 
quantity is shown in Figure \ref{fig:soiw}. 

We also provide a simple fit to the numerical result: 
\be
\beta^\star (w) = \beta^\star_0 \, 
\bigl[ 1 + a (1+w) + b (1+w)^2 + c (1+w)^3 \bigr] \, , 
\label{eq:fit1} 
\ee
where $\beta^\star_0$ is the value for a cosmological constant 
($\beta^\star_0 \approx$ 5.3; \S 3.2) and where the coefficients 
are given by 
\be
a = -2.33 , \qquad \qquad 
b = 2.58 , \qquad \qquad 
c = -1.20 \, . 
\label{eq:fit2} 
\ee 
This simple cubic fit reproduces the numerical results with an
absolute error bounded by $0.035 \beta^\star_0 \approx$ 0.19 (a
relative error of a few percent -- see Figure \ref{fig:soiw}). A
better fit could be obtained by using polynomials of higher order, but
this level of accuracy should be adequate for most applications. With
this fitting polynomial, the sphere of influence for existing
structures, $r_G$ = $(2 G \mob / \beta^\star H_0^2)^{1/3}$, can be
written in the form 
\be
r_G \approx \, 0.7 \, {\rm Mpc} \, \Bigl( 
{\mob \over 10^{12} M_\odot} \Bigr)^{1/3} h_{70}^{-2/3} 
\bigl[ 1 + a (1+w) + b (1+w)^2 + c (1+w)^3 \bigr]^{-1/3} \, . 
\label{eq:fit3} 
\ee

For accelerating universes, we can find the maximum distance 
that a light signal can propagate between now (the present epoch) 
and temporal infinity. This maximum distance $\rmax$ is given by 
\be
\rmax \equiv \int_{t_0}^\infty {c dt \over a(t)} \, = \, 
{c \over H_0} \int_1^\infty {da \over a^2} \Bigl[ 
\om a^{-3} + \vac a^{-p} \Bigr]^{-1/2} \, . 
\ee 
By solving the integral numerically and fitting the result, 
we can write the distance scale in the form 
\be 
\rmax H_0 / c \equiv I(w) \approx  \, I_0 \, 
\bigl[ 1 + {\tilde a} (1+w) + {\tilde b} (1+w)^2 
+ {\tilde c} (1+w)^3 \bigr] \, ,   
\label{eq:rmax} 
\ee 
where $I_0$ = 1.141, $\tilde a$ = 3.073, $\tilde b$ = --12.39, and
$\tilde c$ = 37.13. This fitting function is valid for the range of
equations of state $-1 \le w \le -1/2$. As $w \to -1/3$, the integral
(and hence $\rmax$) becomes divergent. The result is shown in Figure
\ref{fig:rmax}.

With the sphere of gravitational influence defined by equation
(\ref{eq:fit3}) and the maximum distance defined by equation
(\ref{eq:rmax}), we can define the isolation time $t_{\rm iso}$ for
structures through the relation $a(t) r_G \ge \rmax$. The asymptotic
form for the scale factor $a(t)$ is given by 
\be
a(t) \approx \Bigl( {p \over 2} \sqrt{\vac} \, H_0 \, t \Bigr)^{2/p} \, , 
\label{eq:aforw} 
\ee 
where $p=3(1+w)$ as before. The isolation time is then given by 
\be
t_{\rm iso} \approx {2 H_0^{-1} \over p \sqrt{\vac} } \Bigl\{ I(w) 
[\beta^\star(w)]^{1/3} \, c \, \bigl[ 2 G \mob H_0 \bigr]^{-1/3} 
\Bigr\}^{p/2} \, . 
\ee 
Notice that this form applies only for values of $p$ strictly greater
than zero ($w > -1$). To properly take the limit $w \to -1$, $p \to 0$,
the function must include additional terms that are neglected in this 
approximation. 

\section{DIMENSIONAL ANALYSIS} 

In this Appendix, we present a dimensional analysis that illustrates
the fundamental results of this paper in simpler terms.  With the
benefit of hindsight, we can conceptually reproduce many of the
results of this paper.

Because the universe is already dominated by its dark vacuum
contribution, the future behavior of the universe is essentially one
of exponential expansion at a well defined rate. This rate is set by
the Hubble constant. For the sake of definiteness, we will use the
asymptotic value of the Hubble constant $H_\infty$ = $H_0 \sqrt{\vac}$
$\approx$ 59 km s$^{-1}$ Mpc$^{-1}$.  This rate also defines the basic
time scale for the problem, i.e., $\tau = H_\infty^{-1}$ = 17 Gyr.

Using square brackets to denote the units of a given quantity (in
terms of length $L$, time $T$, and mass $M$), we can list the
variables that describe the the asymptotic universe via
\be
[H_\infty] = T^{-1} \qquad [G] = L^3 / M T^2 \qquad [c] = L/T \, . 
\label{eq:basic} 
\ee
These variables can be combined to produce a dimensionless field 
$\Pi_0$ if only if a mass scale $M_0$ is introduced. The field 
$\Pi_0$ is then given by 
\be
\Pi_0 \equiv { G M_0 H_\infty \over c^3 } \, . 
\ee
If no additional variables are introduced into the problem -- no 
additional entities are introduced into the universe -- then 
typically $\Pi_0 \approx 1$, which in turn defines a mass scale 
for the universe. Inserting numerical values, we find $M_0$ $\sim$
$10^{23} M_\odot$ (essentially the same result as that of equation 
[\ref{eq:hormass}]).  

Both our physical intuition and the results of our numerical
simulations indicate that cosmic structure becomes frozen and bound
astronomical objects -- galaxies and clusters -- grow isolated in the
long term. The presence of a cluster or galaxy introduces another 
variable into the problem, namely the mass scale $\mob$. This scale, 
in turn, defines another dimensionless field $\Pi_1$ given by 
\be
\Pi_1 \equiv { G \mob H_\infty \over c^3 } \, \sim 10^{-11} \, , 
\ee
an incredibly small number. This quantity represents the overdensity
of an isolated cluster embedded in an island universe at late times. 

Next, we want to define a length scale $r_0$ associated with the
galaxy or cluster itself. Such a length scale can be defined in several 
ways. The natural length scale of the universe is given by 
$c/H_\infty$ = $r_H$. Since the ratio $r_0/r_H$ is dimensionless, 
we expect that 
\be
r_0 = r_H \Pi_1^n = {c \over H_\infty} \Bigl( 
{G \mob H_\infty \over c^3} \Bigr)^n \, , 
\label{eq:rscale} 
\ee
where the power-law index $n$ is to be determined. If we argue 
that the length scale associated with the galaxy or cluster 
should be non-relativistic, then $n$ must be chosen so that 
the scale $r_0$ does not depend on the speed of light. This 
constraint specifies $n=1/3$ and hence 
\be
r_0 = (G \mob / H_0^2)^{1/3} \, , 
\ee
which is the same as the gravitational sphere of influence defined by
equation (\ref{eq:bound}) (up to dimensionless factors of order
unity). Equation (\ref{eq:rscale}) allows for a second ``natural''
length scale -- that determined by eliminating the Hubble parameter by
using $n = 1$. This choice results in the scale $r = G \mob / c^2$,
which is the length scale that determines the form of the functions
$A(r)$ and $B(r)$ appearing in the metric (eq. [\ref{eq:linein}]).

Another result of this investigation is the time scales for which
objects become isolated and radiation is stretched ``beyond the
horizon''. At this level of analysis, {\it all} of these time scales
are the same and are determined by the asymptotic e-folding time $t =
H_\infty^{-1} \approx$ 17 Gyr. The more detailed mathematical analysis
of the paper includes logarithmic correction factors, as listed in
Table 1. For cosmological events, however, even the longest time scale
is only 1260 Gyr or 74 e-folding times (it also turns out that
$\ln[r_H/\lambda]\sim 75$ -- for `typical' astrophysical photons of
wavelength $\lambda \sim 1 \mu$m). For a universe with a cosmological
constant, the basic result is that all future cosmological events must
unfold with ``nearly'' the same time scale, given by
$H_\infty^{-1}$. For comparison, stellar evolution time scales are
determined by more complicated physics and span a wider range of time
scales, both much shorter ($\sim10^5$ yr for star formation events;
Adams \& Fatuzzo 1996) and much longer ($\sim10^{13}
- 10^{14}$ yr for the duration of the longest-lived stars; AL97,
LBA97). Galactic evolution -- dynamical relaxation and evaporation --
takes place over still longer times ($\sim 10^{20}$ yr; Dyson 1979,
Binney \& Tremaine 1987, AL97).

\end{document}